\documentclass[aps,twocolumn,prd,superscriptaddress,nofootinbib]{revtex4-2}

\usepackage[utf8]{inputenc}

\usepackage{mathtools}
\usepackage{amsfonts}
\usepackage{amssymb}
\usepackage{mathrsfs}
\usepackage{bbm}
\usepackage{slashed}
\usepackage{dsfont}
\usepackage{bm}

\usepackage{graphicx}
\usepackage{subcaption}
\usepackage{color}
\usepackage{array}

\usepackage{placeins}
\usepackage{booktabs}
\usepackage{caption}
\usepackage{float}

\usepackage{xspace}
\usepackage{hyperref}
\usepackage[nameinlink]{cleveref}
\usepackage{bookmark}

\usepackage{xifthen}
\usepackage{xcolor}
\hypersetup{
	colorlinks,
	linkcolor={red!75!black},
	citecolor={blue!75!black},
	urlcolor={blue!75!black}
}
\setkeys{Gin}{width=0.48\textwidth}
\newcommand{\vect}[1]{\bm{#1}}

\usepackage{booktabs}
\usepackage{multirow}

\newcolumntype{C}{>{$}c<{$}}
\AtBeginDocument{
	\heavyrulewidth=.08em
	\lightrulewidth=.05em
	\cmidrulewidth=.03em
	\belowrulesep=.65ex
	\belowbottomsep=0pt
	\aboverulesep=.4ex
	\abovetopsep=0pt
	\cmidrulesep=\doublerulesep
	\cmidrulekern=.5em
	\defaultaddspace=.5em
}

\newcommand{\imag}{\text{i}}

\newcommand{\tinytext}[1]{\text{\tiny{#1}}}

\graphicspath{
	{figures/}
}

\usepackage{xifthen}
\usepackage{xcolor}

\newcommand{\gettitle}{Bound states from the spectral Bethe-Salpeter equation}

\hypersetup{
	colorlinks,
	linkcolor={red!75!black},
	citecolor={blue!75!black},
	urlcolor={blue!75!black},
	pdftitle={\gettitle},
	pdfauthor={Eichmann,Gomez,Horak,Pawlowski,Wessely,Wink},
	pdfkeywords={analytic continuation}
	{correlations functions} {bound states}
	{dyson schwinger equation}
	{real time} {spectral function} {bethe-salpeter equation}
	bookmarksopen=true,
	bookmarksopenlevel=2,
	bookmarksnumbered=true
}

\begin{document}
	
	\title{\gettitle}
	
	\author{Gernot Eichmann}
	\affiliation{Institute of Physics, University of Graz, NAWI Graz, Universit\"atsplatz 5, 8010 Graz, Austria}
	
	\author{Andr\'{e}s~G\'{o}mez}
	\affiliation{Institut f\"ur Theoretische Physik, Universit\"at Heidelberg, Philosophenweg 16, 69120
		Heidelberg, Germany}
	
	\author{Jan~Horak}
	\affiliation{Institut f\"ur Theoretische Physik, Universit\"at Heidelberg, Philosophenweg 16, 69120
		Heidelberg, Germany}
	
	\author{Jan M.~Pawlowski}
	\affiliation{Institut f\"ur Theoretische Physik, Universit\"at Heidelberg, Philosophenweg 16, 69120
		Heidelberg, Germany}
	\affiliation{ExtreMe Matter Institute EMMI, GSI, Planckstr. 1, D-64291 Darmstadt, Germany}
	
	\author{Jonas~Wessely}
	\affiliation{Institut f\"ur Theoretische Physik, Universit\"at Heidelberg, Philosophenweg 16, 69120
		Heidelberg, Germany}
	
	\author{Nicolas~Wink}
	\affiliation{Institut f\"ur Kernphysik (Theoriezentrum), Technische Universit\"at Darmstadt,
		D-64289 Darmstadt, Germany}
	
%%%%%%%%%%%%%%%%%%%%%%%%%%%%%%%%%%%%%%%%%%%%%%%%%%%%%%%%%%
	\begin{abstract}
		We compute the bound state properties of three-dimensional scalar $\phi^4$ theory in the broken phase.
        To this end, we extend the recently developed technique of spectral Dyson-Schwinger equations 
        to solve the Bethe-Salpeter equation and determine the bound state spectrum.
        We employ consistent truncations for the two-, three- and four-point functions of the theory
        that recover the scaling properties in the infinite coupling limit.
        Our result for the mass of the lowest-lying bound state in this limit agrees very well with lattice determinations.
	\end{abstract}
	
	\maketitle

%%%%%%%%%%%%%%%%%%%%%%%%%%%%%%%
\section{Introduction}

The study of bound states with functional methods requires the resummation of a large set of diagrams in a non-perturbative manner.
The standard tool for computing such properties in continuum formulations of quantum field theory (QFT) is the Bethe-Salpeter equation (BSE)~\cite{Salpeter:1951zz,Salpeter:1951sz}. 
Direct extraction of the physical spectrum in terms of the corresponding poles and cuts requires to solve the BSE and, in consequence, knowing the input correlation functions in the timelike domain. This entails additional computational complexity in comparison to calculations in the spacelike domain.

This intricacy has been treated within different approaches. 
Important examples are direct calculations in the complex momentum plane below the onset of singularities~\cite{Maris:1997tm, Maris:1999nt, Alkofer:2002bp, Windisch:2016iud, Eichmann:2016yit},
Cauchy integration~\cite{Fischer:2005en, Krassnigg:2008bob, Dorkin:2013rsa, Dorkin:2014lxa, Rojas:2014aka, Hilger:2017jti} and contour deformation techniques~\cite{Maris:1995ns, Eichmann:2009zx, Strauss:2012dg, Windisch:2012sz, Windisch:2012zd, Pawlowski:2015mia, Pawlowski:2017gxj, Weil:2017knt, Bluhm:2018qkf, Williams:2018adr, Eichmann:2019dts, Miramontes:2019mco, Frederico:2019noo, Santowsky:2020pwd, Eichmann:2021vnj, Huber:2022nzs}, or the Nakanishi method~\cite{Nakanishi:1969ph, Kusaka:1995za, Sauli:2001we, Karmanov:2005nv, Frederico:2013vga, dePaula:2016oct}.
Other works employ reconstructions from Euclidean space data, for example with Padé approximants or the Schlessinger-point method 
\cite{Schlessinger:1968, Rose:2016wqz, Tripolt:2017pzb, Haritan:2017vvv, Tripolt:2018xeo, Alkofer:2018guy, Eichmann:2019dts, Binosi:2019ecz, Santowsky:2020pwd, Huber:2020ngt, Cui:2021vgm, Fukushima:2023wnl}, or ML-inspired reconstructions 
\cite{Cyrol:2018xeq, Kades:2019wtd, Windisch:2019byg, Horak:2021syv, Windisch:2021mem, Pawlowski:2022zhh, Horak:2023xfb,Rothkopf:2022ctl,Rothkopf:2022fyo, Lechien:2022ieg, Lupo:2022nuj, Horak:2023xfb}. These methods have been successful in extracting physical spectra,
but do not fully recover the analytic structure of correlation functions. 

In this work, we introduce the spectral BSE approach, allowing for an efficient solution in the timelike domain by making use of spectral representations for the input correlation functions. 
Their corresponding spectral functions are accessible via the recently developed spectral functional approach~\cite{Horak:2020eng}, which has found application to QCD in the context of DSEs~\cite{Solis:2019fzm,Horak:2021pfr,Horak:2021syv,Horak:2022myj,Horak:2022aza}, and was extended to the functional renormalisation group in~\cite{Braun:2022mgx}, with applications to scalar theories~\cite{Horak:2023hkp} and gravity~\cite{Fehre:2021eob}.
This enables the direct computation of physical masses of bound states and resonances from the corresponding spectral BSE, while also opening the door
to investigating the analytic structure of Bethe-Salpether wave functions~\cite{Maris:1997tm,Maris:1999nt,Kusaka:1997xd,Sauli:2001we,Carbonell:2010zw,Frederico:2013vga,Carbonell:2014dwa,Eichmann:2016yit,Leitao:2017esb,Leitao:2017mlx,dePaula:2017ikc,Biernat:2020yqs,Ydrefors:2019jvu}.

The spectral BSE is set up at the example of a scalar $\phi^4$ theory in three spacetime dimensions. The theory exhibits a second order phase transition and belongs to the Ising model universality class. In the vicinity of the phase transition, the emergence of a two-particle bound state with mass $M \sim 1.8 m$, where $m$ is the mass gap of the theory, has been observed in several works~\cite{Caselle:2001im,Caselle:1999yr,Rose:2016wqz,Nishiyama:2014mra,PhysRevE.77.051112,Lee:2000xna}. The aim of the present study is to approach this bound state from the symmetry-broken phase by considering the infinite coupling limit $\lambda \to \infty$.

\begin{figure*}
    \centering
    \includegraphics[width=.9\textwidth]{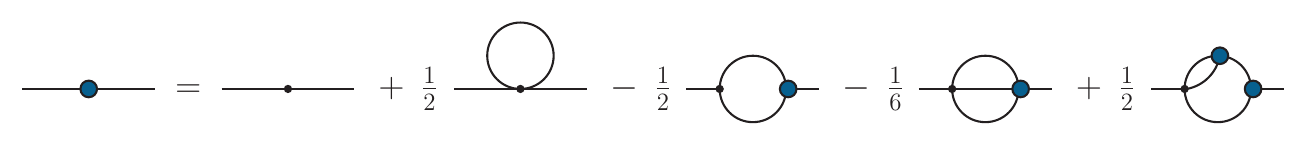} \\[2mm]
    \includegraphics[width=.9\textwidth]{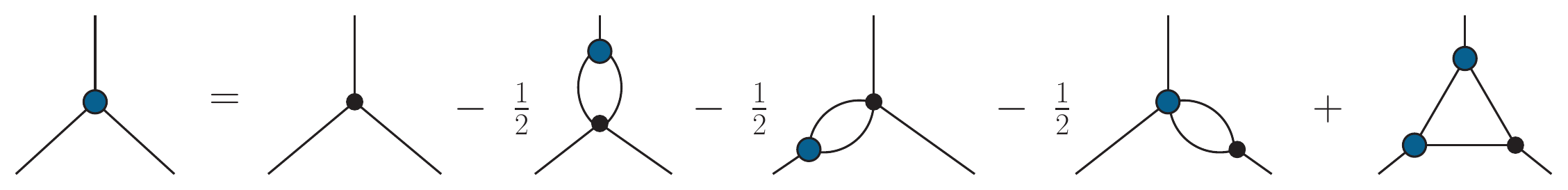}
    \caption{Dyson-Schwinger equations for the two- and three-point functions;
    the latter contains further two-loop terms which are not shown. The full  propagator is represented by a simple line, 
    classical vertices are represented by small black dots and full vertices by large  blue dots, see also \Cref{fig:Notation}.  \hspace*{\fill}}
\label{fig:DSE}
\end{figure*}
\begin{figure}[b]
	\centering
	\subfloat{\includegraphics[width=\linewidth]{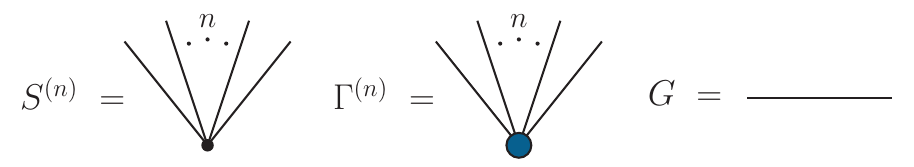}}
	\caption{Diagrammatic notation used throughout this work: small black dots stand for classical vertices, larger blue dots stand for full vertices and lines stand for full propagators.  \hspace*{\fill}}
\label{fig:Notation}
\end{figure}
Our work is outlined as follows. In \Cref{sec:System_overview} we set up the spectral BSE-DSE system for the scalar $\phi^4$ theory and discuss the suitable truncations for the infinite coupling limit.
We present our numerical results for the correlation functions and the bound state position in \Cref{sec:Results}, and conclude in \Cref{sec:conclusion}.
Details on the spectral DSE, the BSE and the numerical implementation can be found in the appendices.

%%%%%%%%%%%%%%%%%%%%%%%%%%%%%%%%%%%%%%%%%%%%%%%%%%%%%%%
\section{Spectral DSEs and BSEs}
\label{sec:System_overview}

In this section, we discuss the spectral BSE-DSE system used in this work. We briefly introduce the spectral DSE approach in \Cref{sec:DSE}, and discuss the employed expansion of our effective potential in \Cref{sec:V_eff}. \Cref{sec:Vertex_approx} is dedicated to a detailed discussion of our systematics relevant for the systematic error control. Finally, in \Cref{sec:full_BSE} we discuss the BSE implementation in the present spectral approach. 

%%%%%%%%%%%%%%%%%%%%%%%%%%%%%%%%%%%%%%%%%%%%%%%%%%%%%%%
\subsection{Dyson-Schwinger equations}
\label{sec:DSE}

The classical action of the scalar $\phi^4$ theory in $d=3$ dimensions reads
\begin{equation}
    S[\varphi] = \int\mathrm{d}^3x \left\{\frac{1}{2} \varphi \left(-\partial^2 + m_0^2 \right) \varphi + \frac{\lambda_\phi}{4!}\varphi^4 \right\} \,,
\label{eq:S_classical}
\end{equation}
where $\lambda_\phi$ is the bare four-point coupling, and $m_0$ is the bare mass of the scalar field. 
Because the coupling constant $\lambda_\phi$ carries a dimension of mass, all quantities can only depend on the dimensionless ratio $\lambda_\phi/m_0$.
In the following, we switch to dimensionless parameters by considering all dimensionful parameters in units of the pole mass $m$. 

The quantum analogue of the classical action~\labelcref{eq:S_classical} is the quantum effective action $\Gamma[\phi]$, see e.g.~\cite{zinn2021quantum}. This is formalised through the master Dyson-Schwinger equation
\begin{equation}
    \frac{\delta\Gamma[\phi]}{\delta \phi} = \bigg\langle \frac{\delta S[\varphi]}{\delta \varphi} \bigg\rangle \,,
\label{eq:DSE}
\end{equation}
stating that the quantum equation of motion of the scalar field $\varphi$ is obtained by varying the quantum effective action w.r.t. the mean field $\phi=\langle \varphi\rangle$. Functional relations for all one-particle irreducible (1PI) correlation functions, 
\begin{equation}
    \Gamma^{(n)}[\phi](p_1, \dots, p_n) = \frac{\delta^n \Gamma[\phi]}{\delta \phi(p_1) \dots \delta \phi(p_n)} \,,
\end{equation}
are obtained from \labelcref{eq:DSE} by the respective $\phi$-derivatives. Generally, the DSE for $\Gamma^{(n)}$ depends on $\Gamma^{(n+2)}$, leading to an infinite tower of coupled equations. A closed system of DSEs is achieved by truncating this tower, e.g., by approximating correlation functions by their classical counterpart from some order $n$ on, $\Gamma^{(m>n)} \approx S^{(m)}$.

The central object in any functional application is the full propagator $G$ which can be obtained from the DSE for its inverse $\Gamma^{(2)}$. The corresponding diagrammatic representation of the latter is shown in the top panel of \Cref{fig:DSE}, containing a tadpole, polarisation, squint and sunset diagram.
Apart from the classical vertices, these diagrams also involve the full three- and four-point vertices (marked by blue blobs). The corresponding DSE for the three-point function is depicted in the bottom panel of \Cref{fig:DSE}.

We employ the spectral DSE framework developed in \cite{Horak:2020eng}. Accordingly, we make use of the K\"all\'en-Lehmann representation for the full propagator,
\begin{equation} 
    G(p) = \int_0^\infty\frac{\mathrm{d} \lambda}{\pi} \frac{\lambda\, \rho(\lambda)}{p^2+\lambda^2} \,,
\label{eq:spec_rep}
\end{equation}
in the diagrams of the gap equation. Within a suitable truncation, the DSE can then be solved directly for timelike momenta due to the resulting perturbative form of the momentum loop integrals. This yields direct access to the spectral function
\begin{equation}
    \rho(\omega) = 2 \, \text{Im} \, G\big(-\imag (\omega + \imag 0^+)\big) \,,
\end{equation}
where we dropped the spatial momentum due to Lorentz covariance.

The spectral function  $\rho$ represents the distribution of the physical states in the full quantum theory, and can be generally parameterised as
\begin{equation}
    \rho(\lambda) = \frac{\pi}{\lambda}\sum_i Z_i\delta(\lambda-m_i)+\tilde{\rho}(\lambda) \,.
\label{eq:specf_decompsoition}
\end{equation}
The $m_i$ are the propagator pole positions for stable one-particle states with residues $Z_i$, whereas
the continuum tail $\tilde{\rho}(\lambda)$ of the scattering states starts at $\lambda = 2m_i$, appearing as a branch cut in the propagator.

In an $s$-channel approximation $p^2=s$ and $t=u=0$, a similar spectral representation can be devised for the four-point function,
\begin{align}
    \Gamma^{(4)}(p) &= \lambda_\phi + \int_0^{\infty}\frac{d\lambda}{\pi} \frac{\lambda\, \rho_4(\lambda)}{p^2+\lambda^2}\,.
\label{eq:4p_spec}
\end{align}
The $\phi^4$ theory in three dimensions is super-renormalisable. The only two superficially divergent diagrams in the propagator DSE are the tadpole and sunset diagrams in \Cref{fig:DSE}, carrying a linear resp. logarithmic divergence. We employ the spectral renormalisation scheme devised in~\cite{Horak:2020eng}. By choosing an on-shell renormalisation condition, the physical scales of our theory are fixed by the pole position of the propagator; see \Cref{app:SpectralDSE} for details.

%%%%%%%%%%%%%%%%%%%%%%%%%%%%%%%%%%%%%%%%%%%%%%%%%%%%%%%
\subsection{Effective potential}
\label{sec:V_eff}
\begin{figure*}[!t]
    \centering
    \includegraphics[width=1.0\textwidth]{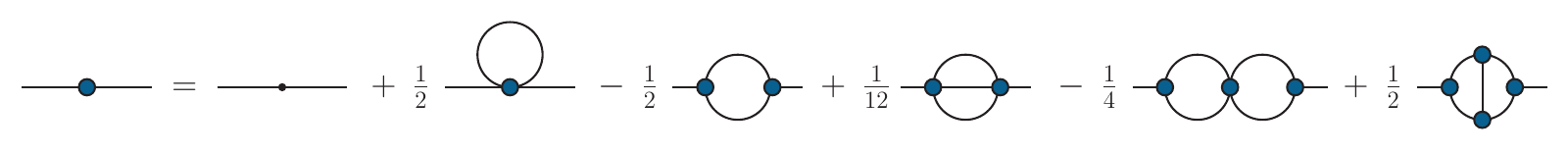}
    \caption{The skeleton expansion for the propagator employed in this work. The full  propagator is represented by a simple line, 
    classical vertices are represented by small black dots and full vertices by large  blue dots, see also \Cref{fig:Notation}.}
\label{fig:Truncations}
\end{figure*}
Instead of resolving the full field dependence of the correlation functions, it is convenient to work on the physical solution to the quantum equation of motion (EoM)
\begin{equation}
    \frac{\delta\Gamma[\phi]}{\delta \phi(x)} = 0 \,.
\label{eq:full_EOM}
\end{equation}
 The symmetry-broken regime of the scalar theory is signalled by a non-vanishing and constant vacuum expectation value $\phi_0=\langle \varphi \rangle$, giving rise to a non-vanishing three-point interaction already at the classical level. For this reason, we solve the Dyson-Schwinger equations in the background
 of the non-vanishing condensate $\phi_0$. The classical vertices $S^{(n)}[\phi_0]$ in the Dyson-Schwinger equations
 are given by 
 \begin{equation}
       S^{(3)}[\phi_0] = \lambda_\phi\,\phi_0\,, \qquad S^{(4)}[\phi_0] = \lambda_\phi\,. 
 \end{equation}

To determine $\phi_0$ in the broken phase dynamically, we expand the
effective potential around the solution of the equation of motion, 
\begin{equation}
    V_{\mathrm{eff}}[\phi] = \sum_{n=2}^{\infty}\frac{v_n}{(2n)!}(\phi^2-\phi_0^2)^n\,.
\label{eq:Veff}
\end{equation}
The $n$-point vertices at vanishing momenta, which we abbreviate by $\Gamma_n$, are then obtained from
\begin{equation}
   \Gamma_n = \Gamma^{(n)}[\phi_0](\boldsymbol{p}_n=0) = \left. \frac{\partial^n V_\mathrm{eff}[\phi]}{(\partial \phi)^n} \right|_{\phi=\phi_0}\,, 
\label{eq:zero-mom-vertices}
\end{equation}
with $\boldsymbol{p}_n=(p_1,...,p_n)$. 
We assume that higher orders in the mean field $\phi_0$ are subleading and truncate the series at second order,
thus parametrising  $V_{\mathrm{eff}}[\phi]$ by its second and third moments $v_2$ and $v_3$. Accordingly, the \mbox{two-}, three- and four-point vertices at zero momentum are given by
\begin{align}\nonumber 
    \Gamma_2 & = \frac13 v_2 \, \phi_0^2\,, \\[1ex]\nonumber 
    \Gamma_3 & = v_2\, \phi_0+\frac{1}{15}v_3 \, \phi_0^3\,, \\[1ex]
    \Gamma_4 & = v_2 +\frac25v_3 \, \phi_0^2\,.
\label{eq:Veff_relations}
\end{align}
By inverting \labelcref{eq:Veff_relations}, one obtains $\phi_0$, $v_2$ and $v_3$ from the zero-momentum correlations functions as
\begin{align}
    \phi_0 &= \frac{3\Gamma_3 - \sqrt{9 \Gamma_3^2-15\Gamma_2 \Gamma_4}}{\Gamma_4}\,,
    \label{eq:phi0_secondO}
\end{align}
and 
\begin{align}
    v_2 &= \frac{6\Gamma_3-\Gamma_4\phi_0}{5\phi_0}\,, \quad
    v_3 = \frac{3(\Gamma_4\phi_0-\Gamma_3)}{\phi_0^3}\,.
\end{align}
The minus sign in front of the square root in \labelcref{eq:phi0_secondO} is determined by the limit $\lambda_\phi/m\longrightarrow0$,
where the full vertices approach their classical values.

%%%%%%%%%%%%%%%%%%%%%%%%%%%%%%%%%%%%%%%%%%%%%%%%%%%%%%%
\subsection{Systematics and truncations}
\label{sec:Vertex_approx}

 Without approximations, the DSE of the inverse two point function carries the full non-perturbative structure of the propagator. In practice, truncations are necessary to deal with the higher correlation functions, which correspond to a certain resummation structure. We are particularly interested in the scaling limit $\lambda_\phi/m\longrightarrow\infty$, where the two-, three- and four-point functions should follow
the scaling relations
\begin{equation} 
    \Gamma^{(2)} \sim p^{2-\eta}\,, \quad
    \Gamma^{(3)} \sim p^{3(1-\eta)/2}\,, \quad
    \Gamma^{(4)} \sim p^{1-2\eta}\,.
\label{eq:scaling}
\end{equation}
The anomalous dimension $\eta$ is known to be $\eta \approx 0.0360$ for the Ising universality class in three dimensions~\cite{Zinn-Justin:1999opn,Gliozzi:2014jsa,Benitez:2011xx}. This imposes tight constraints on the approximation scheme. To ensure the correct scaling behaviour of the diagrams, we employ a skeleton expansion of the propagator DSE. To that end, we convert the classical three- and four-point vertices into full ones.
This procedure introduces additional diagrams which have to be subtracted to remain consistent at a given loop-order. For simplicity, we truncate the expansion of the DSE at two-loop order, leading to the DSE in the skeleton expansion depicted in \Cref{fig:Truncations}. 
Note the changed prefactor of the sunset diagram, stemming from the additional contributions to the tadpole diagram with a full four-point function. The squint diagram is fully absorbed in the now fully dressed polarisation diagram. The latter also produces the double polarisation and the kite, which have to be subtracted. 
Both of them will be ignored in the present work since the kite corresponds to higher-order terms in $\phi_0$ and the double polarisation does not add qualitatively to the analytic structure of the propagator DSE.

\begin{figure}[t]
    \centering
    \includegraphics[width=.9\linewidth]{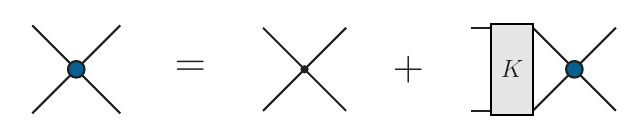}
    \caption{Inhomogeneous BSE for the four-point vertex. The BSE kernel is denoted with a grey box, for the full propagators and vertices see \Cref{fig:Notation}.  \hspace*{\fill}}
\label{fig:InhBSE}
\end{figure}

To close our approximation, we need to specify the higher order correlation functions. We generally perform a zero momentum vertex approximation for all dressed vertices. Nevertheless, since we have argued that the tadpole produces also the sunset topology, we have to include the relevant momentum structure of the four-point function in this diagram. For simplicity, we start from an inhomogeneous BSE, which is shown in \Cref{fig:InhBSE} and reads
\begin{align}\nonumber 
    \Gamma^{(4)}(q_1,q_2,p) & = \lambda_\phi \\[1ex]   % K(q_1,q_2,p)
    &\hspace{-1.5cm}+\int_k K(q_1,k,p)\,G(k_+)\,G(k_-)\,\Gamma^{(4)}(k,q_2,p)\,.
\label{eq:inh_BSE}
\end{align}
Here, $p$ is the total momentum, $q_1$ and $q_2$ are relative momenta, 
$K$ is the two-particle interaction kernel, $G$ is the full propagator with $k_\pm = k\pm p/2$, and $\int_k = \int d^3k/(2\pi)^3$. If we retain only the classical vertex in the kernel, $K(q_1,k,p)=-\lambda_\phi/2$, the equation amounts to a bubble resummation in the $s$-channel approximation,
\begin{align}
    \Gamma^{(4)}(p) = \frac{\lambda_\phi}{1+\lambda_\phi\,\Pi_{\tinytext{fish}}(p)}\,. 
\label{eq:4p}
\end{align}
\Cref{eq:4p} is also readily derived from the DSE of the four-point function in the $s$-channel approximation  with $\phi=0$, also dropping the two-loop terms in the DSE. 

The structure of the `fish diagram' in \labelcref{eq:4p} is the same as that of the polarisation diagram and reads
\begin{equation}
   \Pi_{\tinytext{fish}}(p) = \frac{1}{2} \int_k G(k_+)\,G(k_-)\, ,
\label{eq:fish}
\end{equation}
it corresponds to the spectral integral \labelcref{eq:spec_int} with  $g_{\tinytext{fish}} = 1/2$ in \Cref{app:SpectralDSE}. 
In the limit $\lambda_\phi/m\to 0$, i.e., for a classical propagator $G(k)=1/(k^2+m^2)$, the integral reduces to
\begin{equation}
   \Pi_{\tinytext{fish}}(p) \to \frac{\arctan\sqrt{x}}{16\pi m\,\sqrt{x}}\,, \qquad \Pi_{\tinytext{fish}}(0) \to  \frac{1}{16\pi m}\,,
\end{equation}
with $x=p^2/(4m^2)$. 
The spectral function of the resummed $s$-channel four-point function is extracted by inverting \labelcref{eq:4p_spec} in analogy to the propagator spectral function \labelcref{eq:spec_im}. In this manner, the tadpole with a dressed four-point vertex can be computed in the form of a polarisation diagram with the insertion of this spectral function.

\begin{figure*}
    \centering
    \includegraphics[width=1\textwidth]{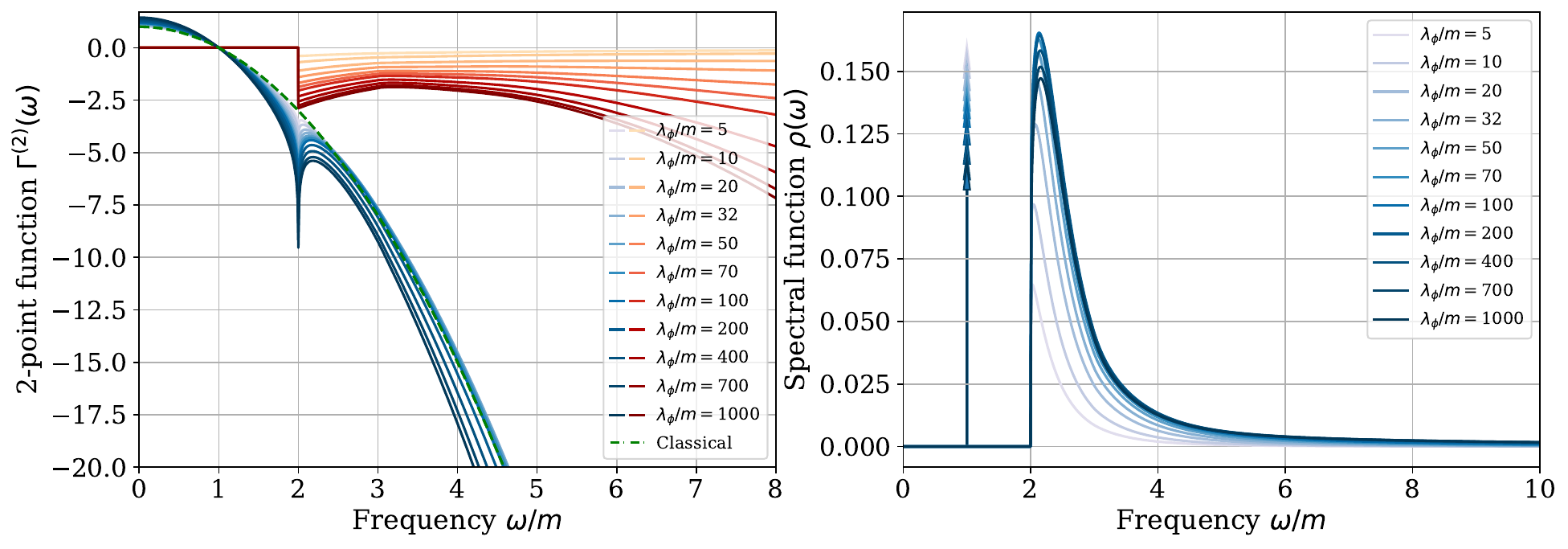}
    \caption{\textit{Left:} Real (blue) and imaginary (red) part of the two-point function obtained in the skeleton expansion, 
    plotted  for real frequencies and different values of $\lambda_\phi/m$. \textit{Right:} Spectral function $\rho(\omega)$. All quantities given in mass units. \hspace*{\fill}}
\label{fig:propagator}
\end{figure*}

For the three-point vertex we consider its DSE up to one-loop terms as shown in \Cref{fig:DSE}.
For simplicity we restrict  ourselves  to vertices at zero momentum, i.e., we assume 
\begin{align}
\Gamma^{(3)}(p_1,p_2,p_3) \approx \Gamma_3\,,\qquad \Gamma^{(4)}(p) \approx \Gamma_4\,.
\end{align}
With the classical three-point function $S^{(3)}[\phi_0]=\lambda_\phi \phi_0$, the DSE reduces to the algebraic equation
\begin{equation}
\begin{split}
    \Gamma_3  = \phi_0 \lambda_\phi &-2\lambda_\phi \Gamma_3 \Pi_{\tinytext{fish}}(0)  -\phi_0 \lambda_\phi \Gamma_4 \Pi_{\tinytext{fish}}(0) \\[1ex]
    & +\phi_0 \lambda_\phi \Gamma_3^2 \,\Pi_{\mathrm{tr}}(0)\,.
\end{split}
\label{eq:3p_DSE}
\end{equation}
Hereby, the triangle diagram
\begin{equation}
  \Pi_{\mathrm{tr}}(p) = \int_k G(k)\,G(k_+)\,G(k_-) 
\label{eq:triangle}
\end{equation}
corresponds to the spectral integral \labelcref{eq:spec_int} in \Cref{app:SpectralDSE} with a prefactor $g_{\mathrm{tr}} = 1$.
For a classical propagator it reduces to
\begin{equation}
   \Pi_{\mathrm{tr}}(p) \to \frac{1}{8\pi m^3} \frac{1}{ x^{3/2}} \arctan\left[ \frac{x^{3/2}}{4+3x}\right]\,,
\end{equation}
with $x=p^2/(4m^2)$ and $\Pi_{\mathrm{tr}}(0) \to 1/(32\pi m^3)$.
One can further eliminate $\phi_0$ by combining \labelcref{eq:3p_DSE,eq:phi0_secondO},
which results in a quartic equation for $\Gamma_3$ and yields
\begin{align}
  (\Gamma_3)^2 &= \frac{5\Gamma_4}{6\lambda_\phi } \frac{6a+b\left(\sqrt{c^2+\frac{12}{5}a}-c\right)}{\Pi_{\mathrm{tr}}(0) \,(2b-5a)}\,, 
 \label{Gamma3-res} 
\end{align}
with the coefficients 
\begin{align}\nonumber 
   a &=\phantom{\frac{2}{5}} \lambda_\phi \Gamma_2 \Pi_{\mathrm{tr}}(0)\,,
   \\[1ex]\nonumber 
   b &= 1 + 2\lambda_\phi \Pi_{\tinytext{fish}}(0)\,, \\[1ex]
   c &= 1 - \frac{2}{5}\lambda_\phi \Pi_{\tinytext{fish}}(0)\,.
\end{align}
This closes our approximation for the DSE system. For the respective results, see \Cref{sec:Results} and especially \Cref{fig:propagator}.
\begin{figure}[t]
    \centering
    \includegraphics[width=.8\linewidth]{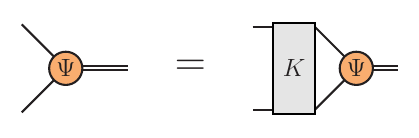}
    \caption{Homogeneous Bethe-Salpeter equation, derived from the inhomogeneous one in \Cref{fig:InhBSE}. The BSE kernel is denoted with a grey box, the BSE wave function with a orange circle and the lines are full propagators.\hspace*{\fill}}
\label{fig:BSE}
\end{figure}
%

%%%%%%%%%%%%%%%%%%%%%%%%%%%%%%%%%%%%%%%%%%%%%%%%%%%%%%%
\subsection{Bethe-Salpeter equation}
\label{sec:full_BSE}

For the calculation of scalar two-particle bound states, we  consider the homogeneous BSE
shown in~\Cref{fig:BSE},
\begin{equation}
    \Psi(q,P) = \int_k K(q,k,P)\,G(k_+)\,G(k_-)\,\Psi(k,P)\,.
\label{eq:BSE_mom}
\end{equation}
Its structure is analogous to that in \Cref{fig:InhBSE} except for the inhomogeneous term:
$\Psi(q,P)$ is the Bethe-Salpeter amplitude, $K$ is the two-particle irreducible kernel,
and $G(k_\pm)$ with $k_\pm = k\pm P/2$ are the dressed propagators.
The total momentum is on-shell, i.e., $P^2 = -M^2$, where $M$ is the mass of the bound state.

Even though the homogeneous and inhomogeneous equations share the same structure,
we note that in our setup they are not directly connected. We employed
\labelcref{eq:inh_BSE} as an ingredient to generate a minimal
four-point vertex that is consistent with scaling, whereas the kernel of the homogeneous BSE is 
related to the self-energy through a functional derivative with respect to the propagator. A possible alternative would be to consider a 4PI system~\cite{Carrington:2004sn,Carrington:2013koa,Carrington:2013jta},
which automatically generates a consistent truncation for the two-, three- and four-point vertices together with the BSE kernel, but this is
beyond the scope of the present work. 
Here we restrict ourselves to the contributions originating from the one-loop terms in the self-energy,
which are the tadpole and  polarisation diagrams. The former generates a scalar four-point vertex
and the latter $t$- and $u$-channel exchanges in the BSE kernel.
Thus, up to two-loop terms the kernel takes the form 
\begin{align} 
    K(q,k,P)  =  \frac{G(q-k)+G(q+k)}{2}\,\Gamma_3^2 - \frac{\Gamma_4}{2}\,,
\label{eq:Kernel_truncation}
\end{align}
where $\Gamma_3$ and $\Gamma_4$ are the three- and four-point vertices at zero momentum.
Note also that the inhomogeneous BSE \labelcref{eq:inh_BSE} does not support
bound states since its kernel $-\lambda_\phi/2$ carries a negative sign,
whereas the additional $t$- and $u$-channel exchanges in the homogeneous BSE change the sign of the kernel to be positive.

Each dressed propagator can then be computed by means of its spectral representation~\labelcref{eq:spec_rep} via the assignment of a unique spectral mass.
Thus, aside from the spectral integrals over $\lambda_1$ and $\lambda_2$ which are performed numerically,
the two internal propagators in the BSE take the form
\begin{equation}
   \frac{1}{k_+^2+\lambda_1^2}\,\frac{1}{k_-^2+\lambda_2^2} = \frac{1}{Q_1^4-Q_2^4}\,,
\label{eq:int_prop}
\end{equation}
and  the sum of the $t$- and $u$-channel contributions in the kernel is
\begin{equation}
    \frac{1}{(q-k)^2+\lambda_3^2}+\frac{1}{(q+k)^2+\lambda_3^2} = \frac{2\,Q_3^2}{Q_3^4-4\,(q\cdot k)^2} \,,
\label{eq:int_ker}
\end{equation}
with
\begin{align}\nonumber 
Q_1^2  &= k^2+P^2/4+(\lambda_1^2+\lambda_2^2)/2\,, \\[1ex]\nonumber 
Q_2^2  &= k\cdot P + (\lambda_1^2-\lambda_2^2)/2\,, \\[1ex]
Q_3^2  &= q^2+k^2+\lambda_3^2\,.
\end{align}
Apart from the spectral representation, we solve the BSE using standard methods, see e.g.~\cite{Eichmann:2016yit,Sanchis-Alepuz:2017jjd}.
By expressing the amplitude $\Psi(q,P)$ in terms of spherical coordinates and
discretising the momentum grid (see \Cref{app:BSE,app:Numerics} for details),
the BSE turns into an eigenvalue equation for a kernel matrix $\mathcal{M}$,
\begin{equation}
   \mathcal{M} \,\Psi_i = \eta_i  \Psi_i \,,
\label{eq:bse}
\end{equation}
whose eigenvalues $\eta_i$ correspond to the ground and excited states and their eigenvectors $\Psi_i$
encode the respective Bethe-Salpeter amplitudes.
Apart from the dependence on the parameter $\lambda_\phi/m$, the eigenvalues depend on the bound state mass ratio $M/m$
through the  total momentum $P$.
We then numerically find the value of $M/m$ for which the relation 
\begin{equation}
    \eta_i\left( \frac{M_i}{m}, \frac{\lambda_\phi}{m}\right) = 1\,,
\label{eq:root}
\end{equation}
holds. This is the on-shell solution for a given state.
The largest eigenvalue corresponds to the ground state, which is the focus of this work.

\begin{figure*}
     \centering
     \includegraphics[width=1\textwidth]{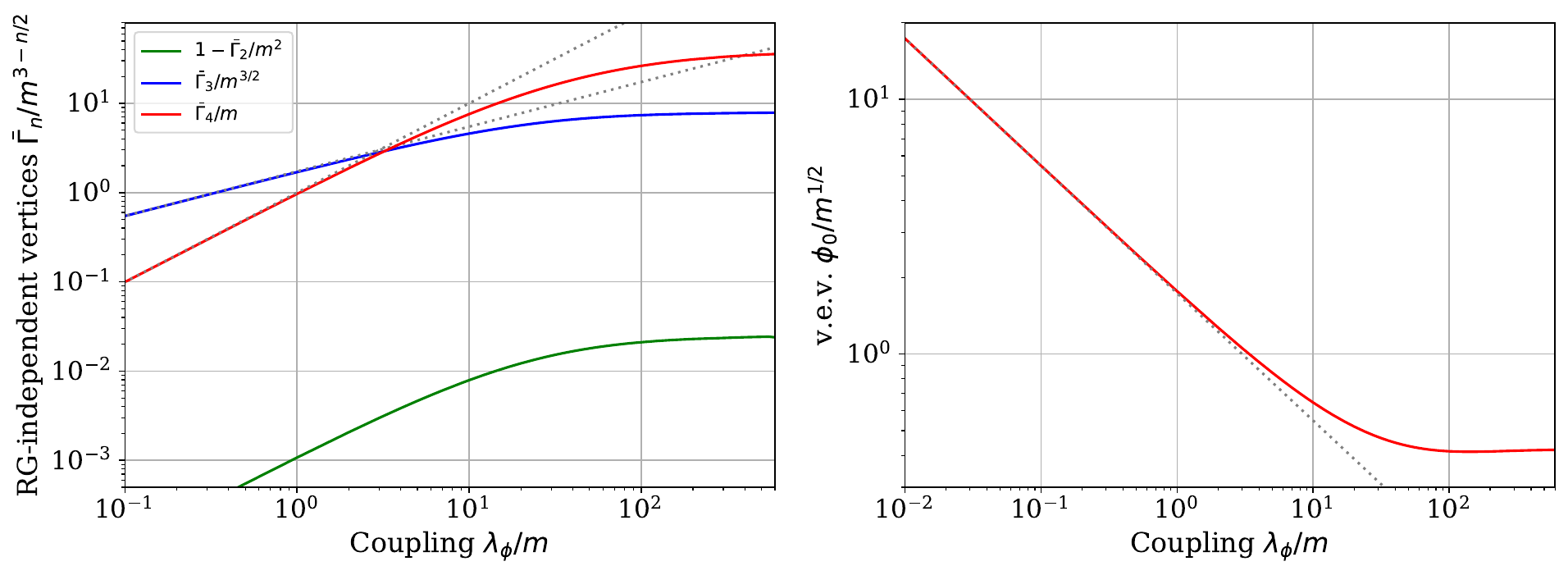}
     \caption{Zero momentum RG-invariant vertices $\bar\Gamma_n/m^{3-n/2}$ with $\bar\Gamma_n$ in \labelcref{eq:RG_vertices}, and of the vacuum expectation value $\phi_0/m^{1/2}$ as functions of the coupling $\lambda_\phi/m$. The dotted gray lines are the tree-level values from \labelcref{eq:vertices-class}. All quantities saturate for large couplings. \hspace*{\fill}}
\label{fig:Mcurv}
 \end{figure*}
%

%%%%%%%%%%%%%%%%%%%%%%%%%%%%%%%%%%%%%%%%%%%%%%%%%%%%%%%
\section{Results}
\label{sec:Results}

Following the discussion of our setup in the previous section, we now present our results.
The left panel of \Cref{fig:propagator} shows the fully dressed inverse propagator for real frequencies $\omega$.
It exhibits an imaginary part, starting at the threshold $\omega = 2m$, which marks the onset of two-particle production.

The right panel of \Cref{fig:propagator} shows respective spectral function $\rho(\omega)$
It is visible how the peak of the spectral tail
increases with the coupling, which implies an increasing dominance of the scattering states.
The peak saturates at around $\lambda_\phi/m \sim 100$  in favor of a broader UV tail. 

The set of RG-independent zero-momentum vertices $\bar{\Gamma}_2$, $\bar{\Gamma}_3$ and $\bar{\Gamma}_4$ are presented in \Cref{fig:Mcurv} as functions of the coupling $\lambda_\phi/m$. These are given by
\begin{equation}
    \bar{\Gamma}_n \equiv \Gamma_n/Z_\phi^{n/2}\,,
\label{eq:RG_vertices}
\end{equation}
where $Z_\phi = 1/Z$ is the wave function renormalisation given at the pole mass as the inverse of the residue. This divides out the RG-running of the external legs, which otherwise leads to a power-law divergence, for further discussions see \Cref{app:ScalingLimit}. 

In the limit $\lambda_\phi/m\to 0$, quantum corrections become negligible and the vertices reduce to their tree-level values
\begin{equation} 
    \frac{\Gamma_2}{m^2} \to 1\,, \quad
    \frac{\Gamma_3}{m^{3/2}} \to  \sqrt{\frac{3\lambda_\phi}{m}}\,, \quad
    \frac{\Gamma_4}{m} \to \frac{\lambda_\phi}{m}\,. 
\label{eq:vertices-class}
\end{equation}
In turn, for asymptotically large couplings $\lambda_\phi/m \to \infty$ we expect a scaling behavior as in this limit we approach the phase transition with $m/\lambda_\phi \to 0$. One can clearly see the deviation from the tree-level behavior
for increasing values of the coupling. With the curvature mass $m_\text{cur}^2 = \bar{\Gamma}_2$, the ratio $m_\text{cur}^2/m^2$ deviates from unity, as is also visible in \Cref{fig:propagator}
 at vanishing momentum. The RG-independent three- and four-point vertices saturate in the large coupling limit. The right panel of \Cref{fig:Mcurv} shows the evolution of the (dimensionless) vacuum condensate $\phi_0/m^{1/2}$,
which starts from its classical result $\sqrt{3m/\lambda_\phi}$ and eventually saturates as well at a non-trivial value. 

\begin{figure}[t]
    \centering
    \includegraphics[width=1\columnwidth]{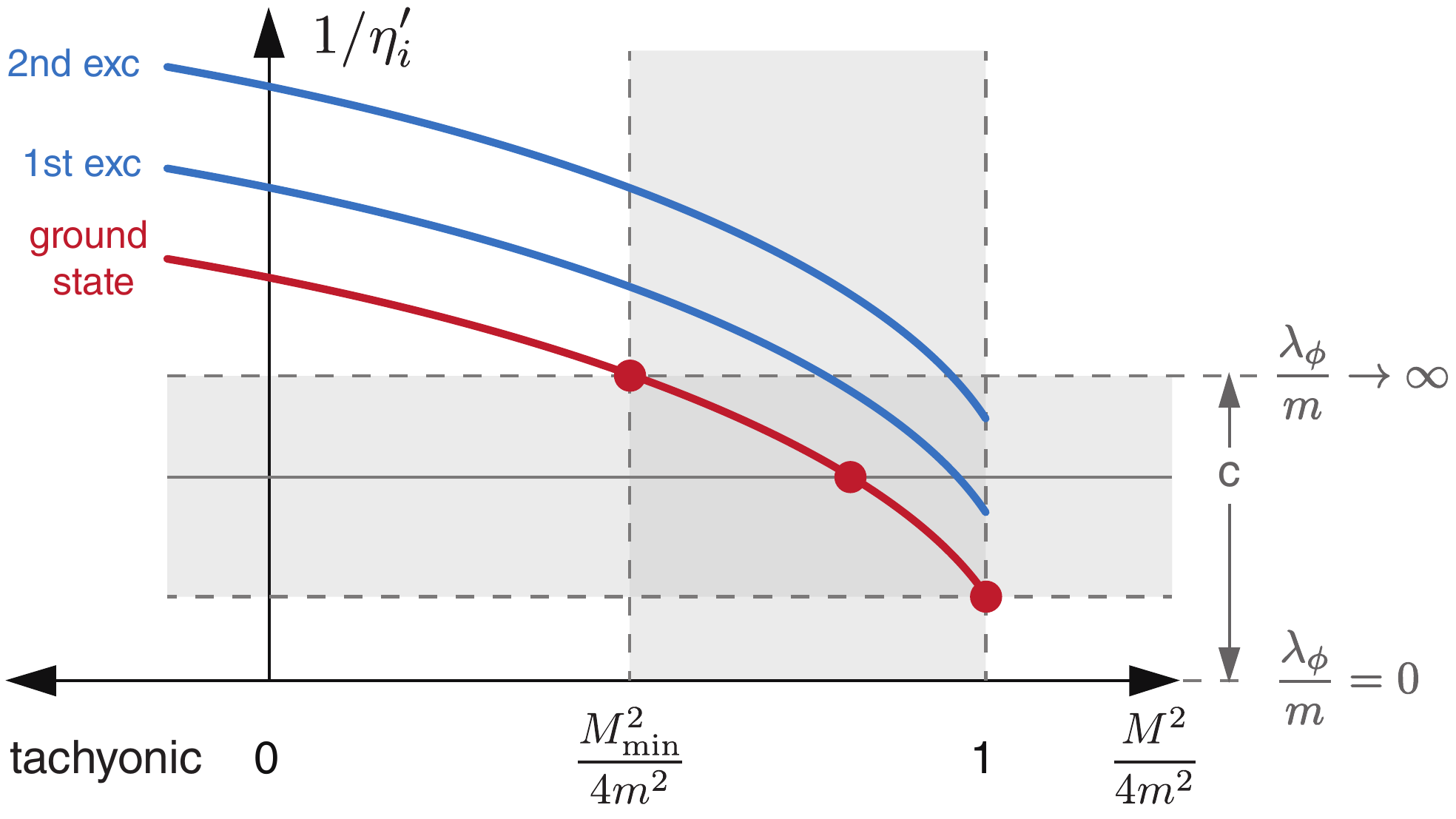}
    \caption{Sketch of the eigenvalue spectrum for a Bethe-Salpeter equation of the form~\labelcref{eq:bse-2}. The masses of the ground and excited states are determined from the condition $1/\eta_i' = c$. If $c$ saturates for $\lambda_\phi/m \to \infty$, the bound state mass has a lower bound $M_\text{min}$. \hspace*{\fill}
             }
\label{fig:evs}
\end{figure}

In summary we find that the dressed RG-invariant vertices calculated from their DSEs eventually saturate, in contrast to their respective tree-level counterparts. This has important consequences for the properties of bound states obtained from the BSE, as it leads to a physical bound-state mass in the scaling limit. To see this, suppose we drop the term $-\Gamma_4/2$ from the BSE kernel~\labelcref{eq:Kernel_truncation}
and solve the BSE with classical (free) propagators only. This yields the massive Wick-Cutkosky model~\cite{Wick:1954eu,Cutkosky:1954ru,Nakanishi:1969ph}
which has been frequently studied in the literature, see e.g.~\cite{Ahlig:1998qf,Kusaka:1995za,Sauli:2001we,Karmanov:2005nv,Frederico:2013vga,Eichmann:2019dts}. If we pull out the dimensionless factor $c=\Gamma_3^2/m^{3}$
from the kernel, the BSE~\labelcref{eq:bse} takes the form 
\begin{equation}
   c\,\mathcal{M}' \,\Psi_i = c \,\eta_i' \, \Psi_i \,. 
\label{eq:bse-2}
\end{equation}
The dimensionless remainder $\mathcal{M}'$ does not depend on $\lambda_\phi/m$ and neither do its eigenvalues $\eta_i'$.
Thus, if we plot the eigenvalue spectrum over the bound state mass, as sketched in \Cref{fig:evs}, the on-shell solution can be read off from the intersection $1/\eta_i' = c$.
The `coupling' $c$ in front of the BSE kernel is now a free parameter that can be tuned arbitrarily; e.g., 
for the tree-level vertex in~\labelcref{eq:vertices-class} it rises linearly with $\lambda_\phi/m$.
In particular, if $c$ is large enough the intersection $1/\eta_i' = c$ occurs at spacelike values  $P^2=-M^2>0$, so that
with increasing coupling one generates tachyonic solutions.

Such a behavior does not happen if the RG-invariant vertices  saturate with $\lambda_\phi/m$, as they do in our system: If the coupling does not exceed
a certain maximum value, then the mass of the bound state is bounded from below and the system cannot become tachyonic.
If in addition the propagators are dressed, as in our case, then the BSE eigenvalues $\eta_i'$ themselves also depend on $\lambda_\phi/m$. Finally,
if we put the four-point vertex back into the kernel there is no longer an overall coupling that can be pulled out since all propagators and vertices
appearing in the BSE kernel are determined from their DSEs.

\begin{figure}
    \centering
    \includegraphics[width=0.49\textwidth]{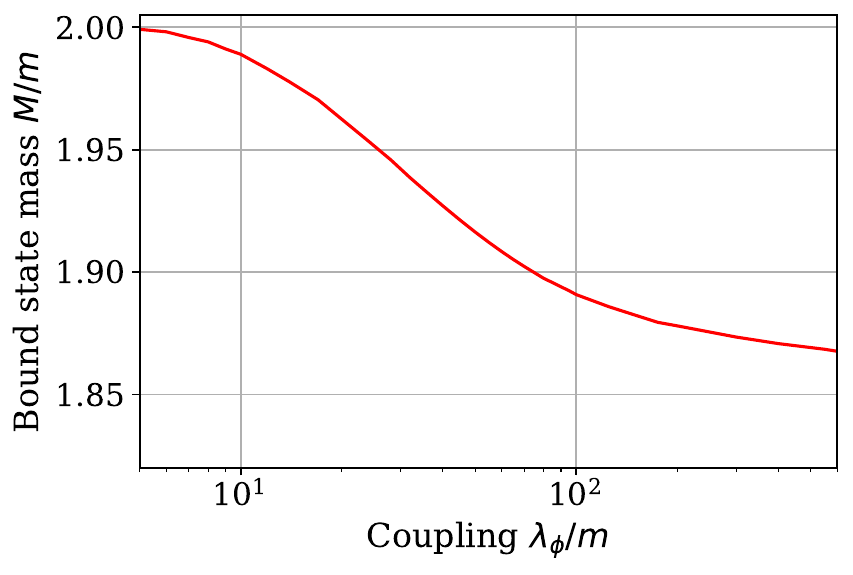}
    \caption{Evolution of the bound state mass $M/m$ as a function of $\lambda_\phi/m$ calculated from the BSE, with propagators and vertices determined from their DSEs.  \hspace*{\fill}}
\label{fig:M_evolution}
\end{figure}

The resulting evolution of the bound state mass ratio $M/m$ with $\lambda_\phi/m$ is shown in \Cref{fig:M_evolution}. 
At $\lambda_\phi/m \approx 5$ the bound state mass is at the threshold $M=2m$. For  smaller couplings 
one might expect either a virtual state like in the massive Wick-Cutkosky model~\cite{Eichmann:2019dts}  or a resonance on the second Riemann sheet.
For larger values of $\lambda_\phi/m$, the bound state mass decreases and eventually saturates. 
Numerical instabilities prohibited us to go beyond $\lambda_\phi/m \gtrsim 10^3$. However, as the bound-state mass already starts to saturate at this scale, an extrapolation allows us to estimate the mass ratio in the scaling limit:
\begin{equation} 
   \frac{M}{m} \approx 1.85 \quad \text{for} \quad \frac{\lambda_\phi}{m} \to \infty\,.
\end{equation}
This is close to the upper range of lattice values $M/m = 1.82(2)$~\cite{Caselle:2001im,Caselle:1999yr,Rose:2016wqz,Nishiyama:2014mra,PhysRevE.77.051112,Lee:2000xna}.
The deviation is of the order of our numerical error of about $1\%$, cf. \Cref{app:Numerics}.
We note again that this result is only possible through a consistent solution for the $n$-point functions,
which underlines the need for systematic truncations with functional methods.

%%%%%%%%%%%%%%%%%%%%%%%%%%%%%%%%%%%%%%%%%%%%%%%%%%%%%%%
\section{Conclusions} 
\label{sec:conclusion}

In this work we studied scalar $\phi^4$ theory in three spacetime dimensions. We determined the mass of the lowest-lying scalar bound state
from its Bethe-Salpeter equation,
whose value in the scaling limit $\lambda_\phi/m\to \infty$ is predicted to be $M/m \approx 1.80 \dots 1.84$ from lattice studies~\cite{Caselle:2001im,Caselle:1999yr,Rose:2016wqz,Nishiyama:2014mra,PhysRevE.77.051112,Lee:2000xna}.
We argued that such a saturation cannot even be achieved qualitatively if the Bethe-Salpeter equation only features tree-level propagators and interactions.
Instead, it requires a consistent truncation of the Dyson-Schwinger equations where not only the propagators but also the vertices acquire a non-perturbative dressing.

To this end, we constructed truncations for the two-, three- and four-point functions where such an internal consistency is explicitly built in.
To solve the system numerically, we employed the spectral DSE approach which allows us to access the timelike behavior of the correlation functions directly.
Up to an anomalous dimension, we find that the three- and four-point vertices saturate in the large coupling limit $\lambda_\phi/m\to \infty$,
and so does the resulting mass of the bound state. Our result $M/m \approx 1.85$ in that limit lies within 1\% of the lattice prediction.

In conclusion, the combination of spectral Dyson-Schwinger and Bethe-Salpeter equations
is a powerful tool that also allows one to access the resonance spectrum above physical thresholds,
or the evolution and melting of bound states with temperature. We hope to report on respective results in the near future.

%%%%%%%%%%%%%%%%%%%%%%%%%%%%%%%%%
\begin{acknowledgements}

We thank Wei-jie Fu and Chuang Huang for discussions. This work is funded by the Deutsche Forschungsgemeinschaft (German Research Foundation) DFG
under Germany’s Excellence Strategy EXC 2181/1 - 390900948 (the Heidelberg STRUCTURES Excellence Cluster) and the Collaborative Research Cen- tre SFB 1225 - 273811115 (ISOQUANT). NW acknowledges support
by the DFG under Project 315477589 – TRR 211
and by the State of Hesse within the Research Cluster ELEMENTS (Project No. 500/10.006).
JH acknowledges support by the Studienstiftung des Deutschen Volkes.
GE acknowledges support from the Portuguese Science Foundation FCT under project
CERN/FIS-PAR/0023/2021 and the FCT computing project 2021.09667.CPCA. \\[2ex]

\end{acknowledgements}

%%%%%%%%%%%%%%%%%%%%%%%%%%%%%%%%%%%%%%%%%%%%%%%%%%%%%%%
\appendix

%%%%%%%%%%%%%%%%%%%%%%%%%%%%%%%%%%%%%%%%%%%%%%%%%%%%%%%
\section{Spectral DSE}
\label{app:SpectralDSE}

In this appendix we provide details on the spectral DSE solution.
Our starting point is the spectral representation \labelcref{eq:spec_rep},
which allows one to compute the full propagator from the spectral function.
This relation can be inverted by analytic continuation to real frequencies, 
\begin{equation}
    \rho(\omega,|\vect{p}|) = 2 \, \mathrm{Im} \, G(-i(\omega+i0^{+}),|\vect{p}|)\,.
\label{eq:spec_im}
\end{equation}
The spectral representation allows one to determine the complete analytic structure of
Feynman diagrams containing full propagators $G_i(p)$, since one only needs to perform
the Euclidean loop integrals $I_j(p,\lambda_i)$ with classical propagators but different spectral masses $\lambda_i$. 
These integrals absorb the full momentum dependence so that  one can perform an analytic continuation into real time.

The self-energy integrals with full propagators are then given by
\begin{equation}
    \Pi_{j}(p) = g_j\prod_{i=1}^{N_j}\bigg(\int_{0}^{\infty}\frac{\mathrm{d}\lambda_i}{\pi} \lambda_i \rho(\lambda_i)\bigg)I_{j}(p,\lambda_i)\,,
\label{eq:spec_int}
\end{equation}
where $\int_0^\infty\mathrm{d}\lambda_i\,\lambda_i/\pi = \int_{\lambda_i}$ 
is the spectral integral and $g_j$ the prefactor of the particular diagram.
In practice, these integrals are performed numerically since also the spectral function is usually computed numerically.
Because the complex structure of the integrand is fully contained within the known functions $I_{j}(p,\lambda_i)$
and only enters in the complex structure of the full diagram $\Pi_{j}(p)$ via its spectral weight $\rho(\lambda_i)$,
these computations are numerically stable.

The computation of real-time Feynman diagrams with full vertices also requires a spectral representation of the latter 
and possesses is own technical limitations~\cite{Wink:2020tnu,Evans:1991ky,Evans:1994cs,Jia:2017niz}.
In this work we ignore the momentum structure of the vertices by approximating them at zero momentum.
The DSE can then be put in the general form 
\begin{equation}
    \Gamma^{(2)}(p) = p^2+m^2+\sum_{j}\Pi_j(p)\,,
\label{eq:Gamma2}
\end{equation}
where $m$ is the bare mass in the classical action and $\Pi_j(p)$ are the spectral integrals~\labelcref{eq:spec_int}
corresponding to the diagrams $I_j(p,\lambda_i)$ with $j=\{\mathrm{tad},\mathrm{pol},\mathrm{squint},\mathrm{sun}\}$, 
whose prefactors $g_j$ come from the combinatorial prefactors in the DSE and the vertices  in the diagrams. 
We importantly remark that these constants are not trivial, as they include the action of the full vertices 
in the diagrams and thus may also depend on the spectral function itself by means of the corresponding DSE of each vertex.

As the full analytical structure of the diagrams $I_j(p,\lambda_i)$ can be computed, the equation can also be represented in real time as
\begin{equation}
    \Gamma^{(2)}(\omega) = -\omega^2+m^2+\sum_{j}\Pi_j(\omega)\,,
\end{equation}
where $\Pi_j(\omega)$ is computed over the analytically continued diagram $I_j(\omega,\lambda_i)$ 
according to \labelcref{eq:spec_im} and carries its own 
real and imaginary component. 

The explicit form of the diagrams is known~\cite{Rajantie:1996np}.
We collect them below alongside their analytical continuation 
\begin{align}
p\to -i(\omega+i0^{+})\,,
\label{eq:p-omega}
\end{align}
where the generated branch cuts  are in accordance with \texttt{Mathematica} conventions. 
We abbreviate $\lambda_{12} = \lambda_1 + \lambda_2$, $\lambda_{123} = \lambda_1 + \lambda_2 + \lambda_3$, etc., 
and we also list the limits at zero momentum, if they are used in the computations:

\medskip \medskip 

\noindent \textbf{Polarisation:}\\[-1ex]

\noindent The polarisation diagram $I_{\mathrm{pol}}$ is given by 
\begin{align}\nonumber 
     I_{\mathrm{pol}}(p,\lambda_1,\lambda_2) &= \frac{1}{4\pi p}\arctan\frac{p}{\lambda_{12}}\,,
 \\[1ex]
     I_{\mathrm{pol}}(0,\lambda_1,\lambda_2) &= \frac{1}{4\pi \lambda_{12}}\,. 
\label{eq:polarisation}
\end{align}
With \labelcref{eq:p-omega} we are led to 
\begin{align}
     I_{\mathrm{pol}}(\omega,\lambda_1,\lambda_2) &= \frac{1}{4\pi \omega}\Bigg[\mathrm{arctanh}\frac{\omega}{\lambda_{12}}+i\arg\bigg(1-\frac{\omega}{\lambda_{12}} \bigg) \Bigg]\,. \nonumber
\end{align}

\noindent \textbf{Sunset:}\\[-1ex]

\noindent The sunset diagram $I_{\mathrm{sun}}$ is given by 
\begin{align}\nonumber 
    & I_{\mathrm{sun}}(p,\lambda_1,\lambda_2,\lambda_3) = \frac{1}{(4\pi)^2}  \\[1ex]
     & \hspace{1.3cm}\times \Bigg[\frac12\ln\frac{1}{\lambda^2_{123}+p^2}-\frac{\lambda_{123}}{p}\arctan\frac{p}{\lambda_{123}}\Bigg]\,,
\end{align}
With \labelcref{eq:p-omega} we are led to 
\begin{align}\nonumber 
    & I_{\mathrm{sun}}(\omega,\lambda_1,\lambda_2,\lambda_3)  = \frac{1}{(4\pi)^2}\Bigg[\frac12\ln\frac{1}{\lambda_{123}^2-\omega^2} \\[1ex]
    & \qquad -\frac{\lambda_{123}}{\omega}\bigg[\mathrm{arctanh}\frac{\omega}{\lambda_{123}}+i\arg\bigg(1-\frac{\omega}{\lambda_{123}}\bigg)\bigg]\Bigg]\,,
\label{eq:sunset}
\end{align}

\noindent \textbf{Squint:}\\[-1ex]

\noindent The squint diagram $I_{\mathrm{squint}}$ is given by 
\begin{align}\nonumber 
    & I_{\mathrm{squint}}(p,\lambda_1,\lambda_2,\lambda_3,\lambda_4)  = \frac{1}{(8\pi)^2\lambda_4\,p}   \nonumber\\[1ex]\nonumber 
    &\qquad \times \Bigg\{ 2\ln\bigg(\frac{\lambda_{234}}{\lambda_{23}-\lambda_4}\bigg)\arctan\frac{p}{\lambda_{14}} \\[1ex]
    &\qquad+i\Bigg[\mathrm{Li}_2\bigg(\frac{ip-\lambda_{14}}{\lambda_{23}-\lambda_4}\bigg) -\mathrm{Li}_2\bigg(\frac{-ip-\lambda_{14}}{\lambda_{23}-\lambda_4}\bigg)  \nonumber \\[1ex]
    &\qquad +\mathrm{Li}_2\bigg(\frac{-ip-\lambda_1+\lambda_4}{\lambda_{234}}\bigg) -\mathrm{Li}_2\bigg(\frac{ip-\lambda_1+\lambda_4}{\lambda_{234}}\bigg) \Bigg]\Bigg\}\,, 
\label{eq:squint}
\end{align}
With \labelcref{eq:p-omega} we are led to 
\begin{align}
 & I_{\mathrm{squint}}(\omega,\lambda_1,\lambda_2,\lambda_3,\lambda_4) = \frac{\mathrm{Re}\,F-i\theta\big(\omega-|\lambda_{14}|\big)\mathrm{Im}\,F}{(8\pi)^2\lambda_4\,\omega}\,, 
 \end{align}
 with
 \begin{align}
    & F  = 2\ln\bigg(\frac{\lambda_{234}}{\lambda_{23}-\lambda_4}\bigg)\mathrm{arctanh}\frac{\omega}{\lambda_{14}} \nonumber \\[1ex]
      & \qquad -\mathrm{Li}_2\bigg(\frac{\omega-\lambda_{14}}{\lambda_{23}-\lambda_4}\bigg) +\mathrm{Li}_2\bigg(\frac{-\omega-\lambda_{14}}{\lambda_{23}-\lambda_4}\bigg) \nonumber \\[1ex]
      & \qquad -\mathrm{Li}_2\bigg(\frac{-\omega-\lambda_1+\lambda_4}{\lambda_{234}}\bigg) +\mathrm{Li}_2\bigg(\frac{\omega-\lambda_1+\lambda_4}{\lambda_{234}}\bigg)\, .
\end{align}

\noindent Finally, the triangle at zero momentum is given by
\begin{equation}
    I_{\mathrm{tr}}(P= p_1= p_2=0,\lambda_1,\lambda_2,\lambda_3) = \frac{1}{4\pi}\frac{1}{\lambda_{12}\,\lambda_{23}\,\lambda_{31}}\,.
\label{app-triangle}
\end{equation}
The tadpole and sunset diagram in the propagator DSE are divergent and need a subtraction. 
We choose an on-shell renormalisation condition $\Gamma^{(2)}(\omega=m_{\mathrm{pole}}) = 0$
such that the renormalised mass is the pole mass $m_{\mathrm{pole}}$. The renormalised DSE thus acquires the form
\begin{equation}
    \Gamma^{(2)}(\omega) = -\omega^2+m_{\mathrm{pole}}^2+\sum_{j}\Big[\Pi_j(\omega)-\Pi_j(m_{\mathrm{pole}})\Big]\,.
\label{eq:ren_spec_DSE}
\end{equation}
We note that no renormalisation of the coupling $\lambda_\phi$ is necessary due to the super-renormalisability 
of  $\phi^4$-theory in three dimensions. Furthermore, the DSE can easily be made dimensionless 
when dividing by $m^2_{\mathrm{pole}}$, thus explicitly recovering the fact that the theory is  determined 
by the dimensionless ratio $\lambda_\phi/m_{\mathrm{pole}}$ only. 
From now on we denote the pole mass by $m$ for simplicity, as also done in the main text.

\begin{figure*}[ht]
    \centering
    \includegraphics[width=1\textwidth]{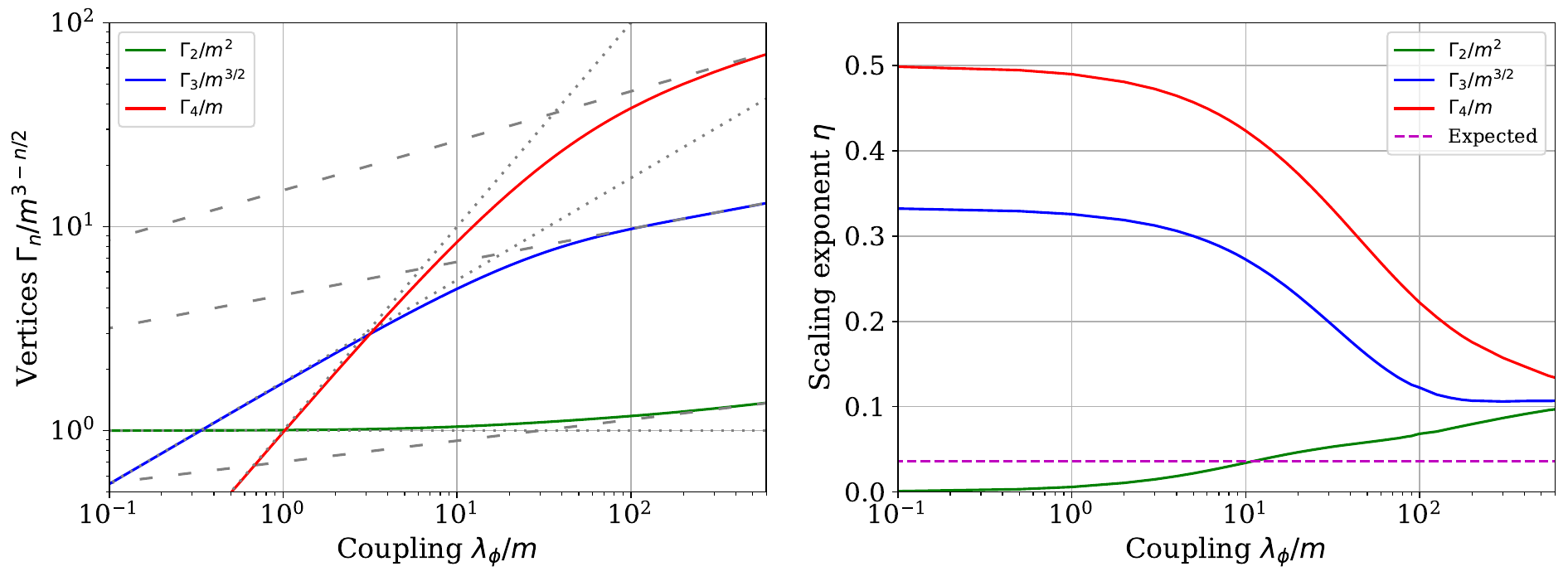}
    \caption{\textit{Left:} Dimensionless zero momentum vertices $\Gamma_n/m^{3-n/2}$ as a function of the coupling $\lambda_\phi/m$. The dotted gray lines are the tree-level values from \labelcref{eq:vertices-class}, whereas the dashed gray lines are the asymptotic curves which follow the scaling limit \labelcref{eq:scaling-2}. \textit{Right:} Scaling exponent of each vertex, extrapolated from \labelcref{eq:scaling-2} for each value of the coupling. All vertices approach the same scaling exponent $\eta \approx 0.11$. The best theoretical prediction of the scaling exponent $\eta = 0.0360$ is represented with the dashed line for comparison.  \hspace*{\fill}}
\label{fig:Vertex_scaling}
\end{figure*}
The spectral DSE \labelcref{eq:ren_spec_DSE} constitutes a non-linear coupled system of integral equations 
for the spectral function $\rho(\lambda)$. 
The spectral integrals on the r.h.s. contain the full vertices, which implicitly 
also depend on $\rho(\lambda)$ through their own DSEs. 
In practice we  solve the spectral DSE by iteration:
We first introduce a reasonable guess for $\rho(\lambda)$ and compute the prefactors
and diagrams according to our truncation scheme. 
We then compute the full propagator via the DSE of the two point function and extract $\rho(\lambda)$ from the l.h.s. of \labelcref{eq:spec_im},
which we introduce again in the spectral DSE.
The three- and four-point vertices along with the condensate $\phi_0$ are computed in parallel.
We repeat the process until convergence is achieved. 

According to the general form \labelcref{eq:specf_decompsoition}, one obtains the mass $m_i$ of 
each stable one-particle state by computing the zeroes of the two-point function and determine their residue as
\begin{equation}
Z_i = -\frac{2m_i}{\partial_\omega \Gamma^{(2)}(\omega)}\bigg|_{\omega=m_i}\,.
\end{equation}
In our case we only find one root coming from the original one-particle state. 
The corresponding residue is bounded by 1, being exactly 1 for the non-interacting theory 
and expected to decrease as dispersive states become more relevant in the interacting theory.
 The continuous tail from these dispersive states can be computed from \labelcref{eq:spec_im}. 
 This tail starts at the two-particle threshold $2m$ and goes to zero in the UV, although 
 it also has successive tails at every subsequent $n$-particle threshold which are suppressed 
 by their corresponding mass.

%%%%%%%%%%%%%%%%%%%%%%%%%%%%%%%%%%%%%%%%%%%%%%%%%%%%%%%
\subsection{Scaling limit}
\label{app:ScalingLimit}

 For the (dimensionless) zero-momentum vertices, the scaling relations suggest %\cAGA{signs of exponents are flipped}
\begin{align} 
     \frac{\Gamma_2}{m^2} \sim \left[\frac{\lambda_\phi}{m}\right]^{\eta} \,, \quad
     \frac{\Gamma_3}{m^{3/2}} \sim \left[\frac{\lambda_\phi}{m}\right]^{\frac{3\eta}{2}} \,, \quad
     \frac{\Gamma_4}{m} \sim \left[\frac{\lambda_\phi}{m}\right]^{2\eta} \,,
\label{eq:scaling-2}
\end{align}
where $\eta$ is the anomalous dimension and $\lambda_\phi/m$ takes the role of the momentum in \labelcref{eq:scaling} in \Cref{sec:Vertex_approx}. With \labelcref{eq:scaling-2} we can infer a scaling exponent $\eta$ from converging results of the RG-variant correlation functions displayed in \Cref{fig:Vertex_scaling}. The left panel shows the set of zero-momentum vertices $\Gamma_2$, $\Gamma_3$ and $\Gamma_4$ as functions of the coupling $\lambda_\phi/m$. These approach their tree level values (dotted lines) for small couplings, whereas for larger couplings they asymptotically approach a scaling behaviour which matches \labelcref{eq:scaling-2}. By applying a logarithmic derivative, we obtain the associated scaling exponent $\eta$. This is shown in the right panel of \Cref{fig:Vertex_scaling}, where it is visible how all three vertices approach a common scaling exponent $\eta \approx 0.11$, which is in agreement with fRG calculations on the Keldysh contour in the broken phase~\cite{Roth:2023wbp}. The authors find a deviation between the broken and symmetric phase which might point towards an interplay of $\eta$ and $\nu$ in the broken phase. However, fRG results in the symmetric phase in a similar truncation point towards a very small scaling region, where momentum scaling of the vertices and in particular the two-point function emerge~\cite{Kockler}, which is by no means reached in the present work.

 %%%%%%%%%%%%%%%%%%%%%%%%%%%%%%%%%%%%%%%%%%%%%%%%%%%%%%%
\subsection{Modified skeleton expansion}
\label{app:ModifiedSkeleton}
\begin{figure*}[t]
    \centering
    \includegraphics[width=1.0\textwidth]{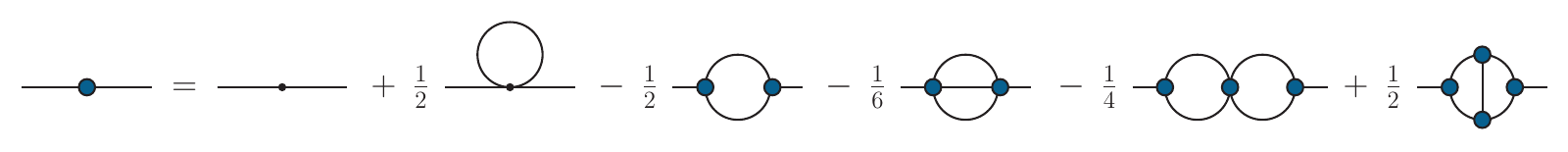}
    \caption{The modified skeleton expansion.}
\label{fig:ModSkeleton}
\end{figure*}
To estimate the relevance of the resummed 4-point function, we devise another approximation scheme, where we drop the full vertex in the tadpole. The  diagrammatic depiction of the gap equation is provided in  \Cref{fig:ModSkeleton}. The re-adjusted prefactor of the sunset diagram guarantees two-loop consistency, and hence both approximations agree at two loop, but not beyond.

\begin{figure*}
     \includegraphics[width=1\textwidth]{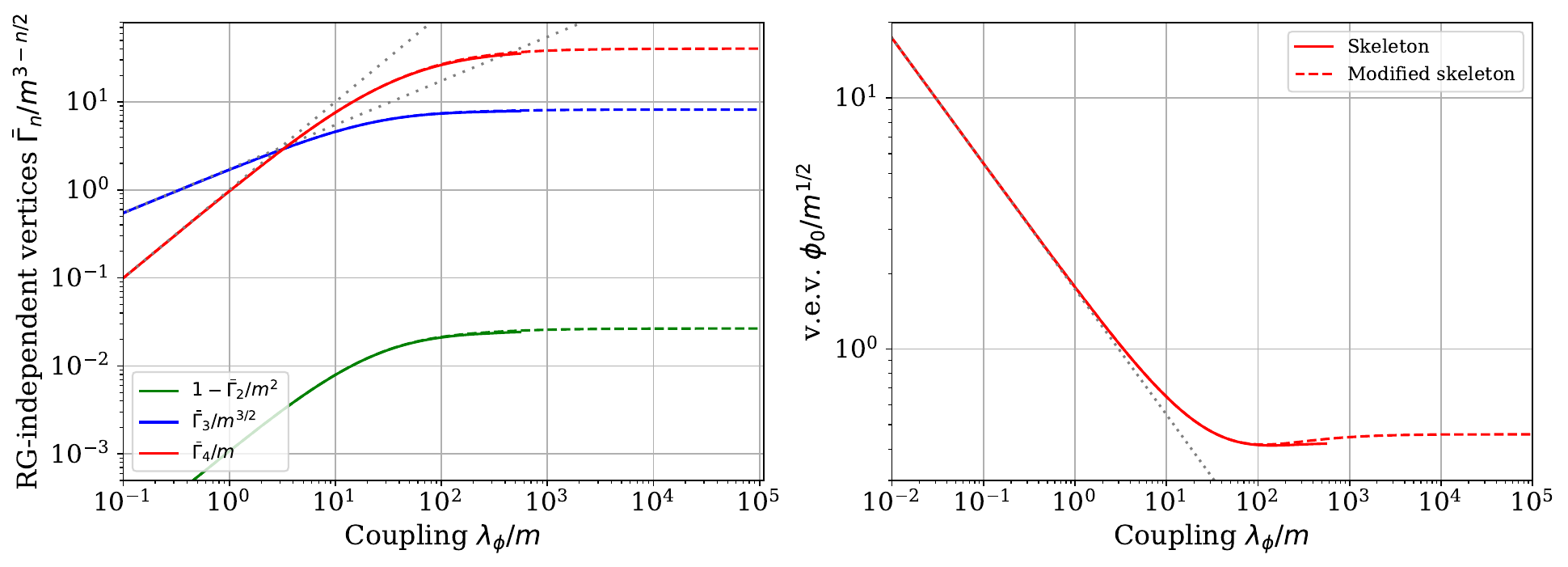}
     \caption{Evolution of the dimensionless zero momentum RG-independent vertices \labelcref{eq:RG_vertices} and of the vacuum expectation value $\phi_0/m^{1/2}$ with the coupling $\lambda_\phi/m$  compared for the two DSE truncations: Solid curves come from the skeleton expansion of the main text, and the dashed lines correspond to the modified skeleton expansion of this appendix. The dotted gray lines are the tree-level values from \labelcref{eq:vertices-class}. All quantities saturate for large couplings. \hspace*{\fill}}
\label{fig:vertices_comparison}
 \end{figure*}

The respective DSE results are presented in \Cref{fig:vertices_comparison}. In comparison to the full results, the modified skeleton approximation does not exhibit scaling. All quantities approach a finite large coupling value, including the residue of the mass pole of the spectral function. The reason for this is that the zero momentum approximation of all of the vertices in this expansion fails to correctly represent the approaching quadratic divergence which is supposed to be present in the scaling limit. This nevertheless means that this DSE system is numerically stable, allowing us to solve for arbitrary values of the coupling.

We see that the full momentum dependence of the tadpole in the skeleton expansion of the main text is the reason for successfully achieving a scaling behaviour. Furthermore, this scaling behaviour is also what produces the numerical instabilities that prohibit us from obtaining solutions for $\lambda_\phi/m \gtrsim 10^3$.

\begin{figure}[b]
    \centering
    \includegraphics[width=0.49\textwidth]{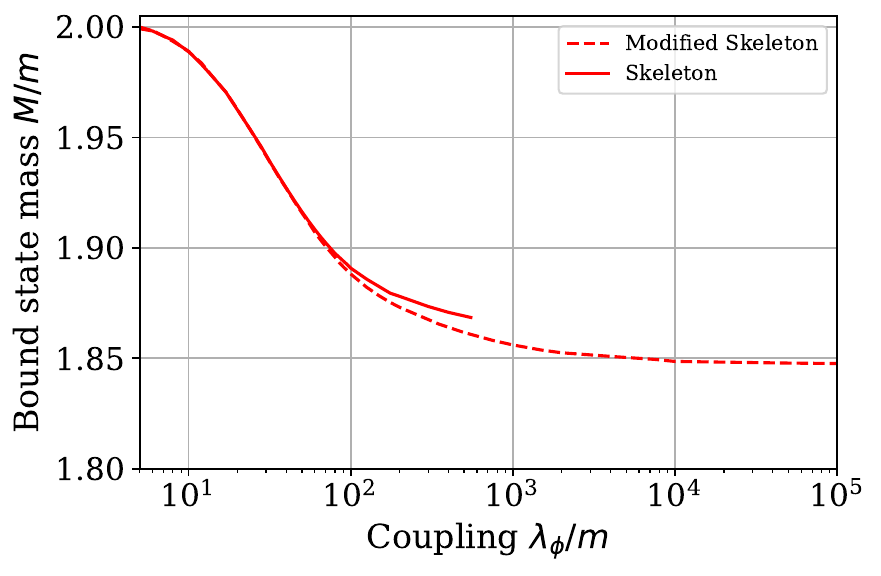}
    \caption{Evolution of the bound state mass $M/m$ as a function of $\lambda_\phi/m$ calculated from the BSE, with propagators and vertices determined from their DSEs.  \hspace*{\fill}}
\label{fig:M_comparison}
\end{figure}

We show the corresponding bound state mass in \Cref{fig:M_comparison} for which we also made use of the scaling Kernel \labelcref{eq:Kernel_truncation}. This was devised in order to compare the differences coming solely from the changes in the self energy of each approximation. We see that, just as with the RG invariant vertices, the bound state mass of both approximations is in very good agreement even when close to the phase transition. This allows us to confidently extrapolate the limiting bound state mass of the skeleton expansion at the given value $M/m \approx 1.85$.

%

%%%%%%%%%%%%%%%%%%%%%%%%%%%%%%%%%%%%%%%%%%%%%%%%%%%%%%%
\subsection{Spectral convolution}
\label{app:SpectralConvolution}
Here we describe a method used to effectively compute spectral integrals even in the large coupling limit where one has an increased weight of the spectral tail. 

Suppose one has a 2-dimensional spectral integral over two spectral functions which are not necessarily the same, 
\begin{equation}
    \Pi_j(\omega) = \int_0^{\infty}\int_0^{\infty}\frac{\mathrm{d}\lambda_1\mathrm{d}\lambda_2}{\pi^2} \lambda_1\lambda_2 \rho_1(\lambda_1)\rho_2(\lambda_2)I_{j}(\omega,\lambda_1,\lambda_2)\,,
\end{equation}
but with a diagram (such as the polarisation diagram) which solely depends on the sum of the spectral weights: 
$I_j(\omega,\lambda_1,\lambda_2) = I_j(\omega,\lambda_1+\lambda_2)$. Then, a change of variables $\eta = \lambda_1+\lambda_2$ 
and a reparametrisation of the region of integration transforms this into a one-dimensional spectral integral 
\begin{equation}
    \Pi_j(\omega) = \int_0^{\infty}\frac{\mathrm{d}\eta}{\pi} \eta \rho_{12}(\eta)I_{j}(\omega,\eta)\,,
\end{equation}
over a new spectral function, which is given by the convolution of the two initial spectral functions, 
\begin{equation}
    \rho_{12}(\eta) = \frac{\pi}{\eta}\int_0^{\eta}\frac{\mathrm{d}\lambda_2}{\pi^2}(\eta-\lambda_2)\lambda_2\rho_1(\eta-\lambda_2)\rho_2(\lambda_2).
\label{eq:spec_conv}
\end{equation}
This transforms a two-dimensional spectral integral into a one-dimensional integral over a convolution of the 
spectral function constituents, which we call spectral convolution. The convolution of spectral functions 
inherits all of the basic properties of the convolution. For a basic decomposition of the form
\begin{equation}
    \rho_i(\lambda_i) = \frac{\pi}{\lambda_i}Z_i\delta(\lambda_i-m_i)+\tilde{\rho}_i(\lambda_i)\,,
\label{eq:basic_decomposition}
\end{equation}
the spectral convolution results in
\begin{align}
    \rho_{12}(\eta) & = \frac{\pi}{\eta}Z_1Z_2\delta(\eta-m_1-m_2)+\frac{\eta-m_1}{\eta}Z_1\tilde{\rho}_2(\eta-m_1)  \nonumber \\[1ex]
    &+\frac{\eta-m_2}{\eta}Z_2\tilde{\rho}_1(\eta-m_2)+\tilde{\rho}_{12}(\eta)\,,
\end{align}
where $\tilde{\rho}_{12}(\eta)$ is the convolution of the tails following  from \labelcref{eq:spec_conv}. 
Thus, the spectral convolution has the same decomposition of the initial spectral functions in a way 
which compactifies all of the information of contributing to the spectral integral of its constituents.

The spectral convolution can be easily generalised for three-dimensional spectral integrals of diagrams 
which depend on the sum of the three spectral weights (such as the sunset) by the repeated convolution 
of spectral functions, or partially implemented in a diagram which depends on the sum of only some 
of the spectral weights (such as the squint).

The advantage of this method is that the convolution of the spectral functions does not depend on the diagrams. 
Thus, it is an integral over smooth functions which can be precomputed and reutilised for any diagram 
whose components also depend on the sum of two spectral weights. But most importantly, it considerably optimises 
the numerical implementation of the spectral integrals, because the resolution of the curve of poles in two dimensions, 
usually given by the equation $\omega = \lambda_1+\lambda_2$, simplifies to the resolution of a pole at a point $\omega = \eta$.

%%%%%%%%%%%%%%%%%%%%%%%%%%%%%%%%%%%%%%%%%%%%%%%%%%%%%%%
\section{Bethe-Salpeter equation}
\label{app:BSE}

Here we give details on the BSE solution discussed in \Cref{sec:full_BSE}. 
According to the spectral decomposition~\labelcref{eq:spec_rep}, the BSE kernel reads 
\begin{align}\nonumber 
    K(q,k,P) &= \Gamma_3^2 \int_0^\infty \frac{d\lambda_3}{\pi}\,\lambda_3\,\rho(\lambda_3)\,\mathcal{K} -  \frac{\Gamma_4}{2}\,, \\[1ex]
    \mathcal{K} &= \frac12\left[\frac{1}{(q-k)^2+\lambda_3^2}+\frac{1}{(q+k)^2+\lambda_3^2}\right]\,,
\end{align}
where $\mathcal{K}$ comes from the $t$- and $u$-channel contributions in the kernel. 

For the explicit coordinate representation we follow the conventions of~\cite{Eichmann:2019dts} 
and express the momenta in three-dimensional Euclidean spherical coordinates, 
\begin{align}\nonumber 
    \frac{q}{m} = & \sqrt{X}\begin{pmatrix} 0 \\ \sqrt{1-Z^2} \\ Z \end{pmatrix}\,,\\[2ex]
    \frac{k}{m} =& \sqrt{x}\begin{pmatrix} \sqrt{1-z^2}\,\sin\varphi \\ \sqrt{1-z^2}\,\cos\varphi \\ z \end{pmatrix}\,,
\end{align}
with $P = 2m\sqrt{t}\,( 0, 0, 1 )$, where at the end of the calculations we take $\sqrt{t} = iM/(2m)$. This implies
\begin{align}
\begin{array}{rl}
    q^2 &= m^2 X\,, \\[1ex]
    k^2 & = m^2 x\,, \\[1ex]
    P^2 &= 4m^2 t = -M^2\,, 
\end{array}\qquad
\begin{array}{rl}
    q\cdot P & = 2m^2\sqrt{Xt}\,Z\,, \\[1ex]
    k\cdot P &= 2m^2\sqrt{xt}z\,, \\[1ex]
    q\cdot k & = m^2\sqrt{xX}\,\Omega, 
\end{array}
\end{align}
with $\Omega = zZ+\sqrt{1-z^2}\sqrt{1-Z^2}\cos\varphi$. The integral measure then takes the form
\begin{equation}
    \int\mathrm{d}^3k = \frac{m^3}{2}\int_0^\infty \mathrm{d}x\sqrt{x}\int_{-1}^{1}\mathrm{d}z\sqrt{1-z^2}\int_0^{2\pi}\mathrm{d}\varphi\,.
\end{equation}
Equation \labelcref{eq:int_prop} turns into 
\begin{align}
\frac{1}{k_+^2+\lambda_1^2}\frac{1}{k_-^2+\lambda_2^2} = \frac{1}{m^4}\frac{1}{\tilde{Q}_1^4-\tilde{Q}_2^4}\,,
\label{eq:int_prop2}
\end{align}
with 
\begin{align}
\begin{split}
\tilde{Q}_1^2 & = x + t +\frac{\lambda_1^2+\lambda_2^2}{2m^2}\,, \\[1ex]
\tilde{Q}_2^2 & = 2\sqrt{xt} z + \frac{\lambda_1^2 - \lambda_2^2}{2m^2}\,,
\end{split}
\end{align}
and the $t$- and $u$-channel exchange kernel $\mathcal{K}$ reads
\begin{equation}
    \mathcal{K} = \frac{1}{m^2}\frac{X+x+\lambda_3^2/m^2}{\big(X+x+\lambda_3^2/m^2\big)^2-4Xx\,\Omega^2}\,.
\label{eq:int_ker2}
\end{equation}
The total bound state momentum $P$ is evaluated in the timelike region, but this analytic continuation is trivial within our 
spectral decomposition. 
Despite the imaginary term $\sim\sqrt{t}$ in the denominator of \labelcref{eq:int_prop2}, the product of the dressed propagators 
is  real because its imaginary part is odd 
in $(\lambda_1,\lambda_2)$ and integrates to zero in the spectral integrals. 
Furthermore, because the 
spectral variables $\lambda_1$ and $\lambda_2$ only take values at $m$ and above $2m$, \labelcref{eq:int_prop2} 
is finite for all masses $M<2m$ below the two-particle threshold. Finally, the integration over 
$\varphi$ in \labelcref{eq:int_ker2} can also be done analytically using
\begin{equation}
    \int_0^{2\pi}\frac{\mathrm{d}\varphi}{1-(a+b\cos\varphi)^2} = \sum_{\chi=\pm1}\frac{\pi}{\sqrt{(1+\chi a)^2-b^2}}\,.
\end{equation}
\begin{figure}
    \centering
    \includegraphics[width=0.49\textwidth]{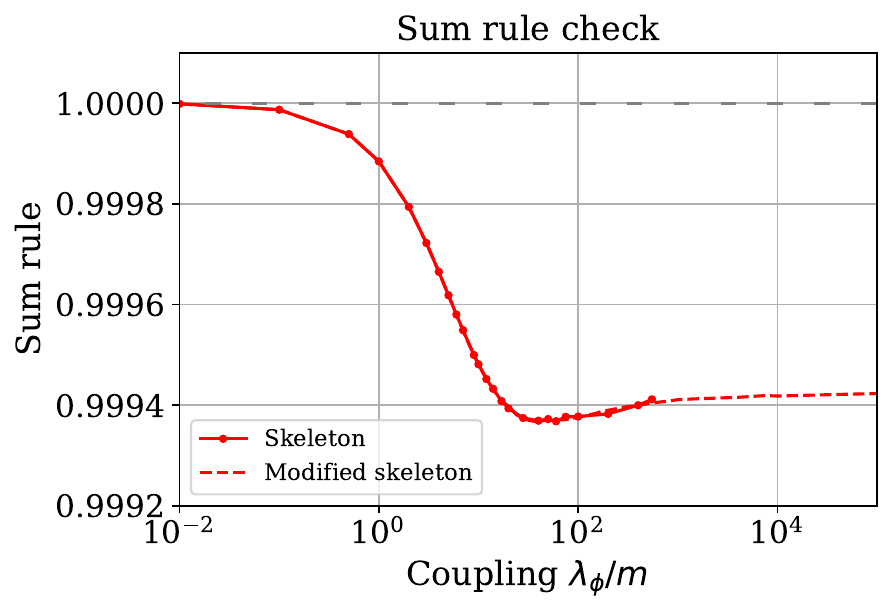}
    \caption{Results for the sum rule (the integral of the spectral function) from each DSE truncation.The deviations from unity are well within our conservative estimate $10^{-2}$ for the relative numerical error. \hspace*{\fill}}
\label{fig:sum_rule}
\end{figure}
%

%%%%%%%%%%%%%%%%%%%%%%%%%%%%%%%%%%%%%%%%%%%%%%%%%%%%%%%
\section{Numerics}
\label{app:Numerics}

Let us finally discuss the numerical details in the DSE and BSE solution.
In every computation, the spectral integrals were performed with an adaptive quadrature routine with error 
equal to $1\times 10^{-8}$ via the spectral convolution method. The diagrams and spectral functions were computed for a finite sampling of points on two intervals: One small interval $(0,b)$ for a medium sbehaviourd $b$ which gave us the momentum features in detail, and another bigger interval $(b,c)$ which gave us the correct weight of the corresponding UV tail. The sampling on the first interval was performed with an adaptive parallel function evaluation algorithm~\cite{Nijholt2019}, whereas the sampling of the UV tail was performed on a logarithmic grid. We generally chose the values $b = 20m$ and $c = 10\lambda_\phi + 1000m$, for which an accurate interpolation of the momentum features and the UV tail was obtained. The only exception was the spectral function of the 4-point function, for which $b = \mathrm{max}(20m,\lambda_\phi/2)$ needed to be taken while still leaving the same value for $c$. For the first interval we chose a sampling of $200$ points for both spectral functions, $120$ for the polarization and $100$ for the sunset and tadpole diagrams. For the second interval we used $100$ points for all of the objects. With the aforementioned sampling, all of the objects were interpolated with a piecewise cubic Hermite interpolating polynomial which both ensures smoothness and a monotonous tail for monotonous data, which is important in the UV with a logarithmic spacing.

The process of iterating the spectral function and vertices back into the DSE was made until 
all of the parameters ($Z$, $m_{\mathrm{cur}}/m$, $\Gamma_4/m$, $\Gamma_3/m^{3/2}$ and $\phi_0/\sqrt{m}$
 had a relative change no greater than $0.2\times 10^{-3}$ between iterations. This leads to our conservative estimate for the numerical error of the order of $10^{-2}$ for our results. 

The  computation of the BSE matrix was performed on a discretised momentum grid of 
$(N_X,N_Z,N_x,N_z)=(40,40,40,40)$ points. The root finding algorithm for solving \labelcref{eq:root} 
was implemented with an accuracy of $1\times 10^{-3}$ for $M/m$. The use of finer grids for the 
BSE matrix did not change the  value of the resulting mass within this level of accuracy. 
Nevertheless, coming from the estimated numerical error of our DSE computations, 
we expect our final results to have an error of $10^{-2}$.

As a consistency check we computed the sum rule (integral of the spectral function) for each value 
of the coupling and each DSE truncation; this is shown in \Cref{fig:sum_rule}. We obtain deviations 
no bigger than $3\times 10^{-3}$ from the theoretical result, which means that the spectral sum rule 
is satisfied within the estimated numerical error of $10^{-2}$.

%%%%%%%%%%%%%%%%%%%%%%%%%%
%\bibliographystyle{apsrev4-2}
\bibliography{references.bib}

\end{document}